# An Efficient Shock Advice Algorithm based on K-Nearest Neighbors for Automated External Defibrillators

Dao Thanh Hai
Posts and Telecommunications
Institute of Technology
Hanoi, Vietnam
haidt102@gmail.com

Nguyen Minh Tuan
Post and Telecommunication
Institute of Technology
Hanoi, Vietnam
nguyenminhtuan80@gmail.com

Nguyen Thi Thu-Hang
Posts and Telecommunications
Institute of Technology
Hanoi, Vietnam
hangntt@ptit.edu.vn

Chau Hai Le
Post and Telecommunication
Institute of Technology
Hanoi, Vietnam
chaulh@ptit.edu.vn

*Abstract*— **Shockable rhythms, namely ventricular fibrillation, and ventricular tachycardia, are the main cause of sudden cardiac arrests (SCA), which can be detected quickly by the automated external defibrillator (AED) devices. In this paper, a simple and efficient shock advice algorithm is developed, and it can be practically applied in AED. The proposed algorithm consists of K-nearest neighbor classifier and an optimal set of 36 features, which are extracted from original ECG and shockable, non-shockable signals using modified variational mode decomposition (MVMD) technique. Cross validation procedure and sequential forward feature selection are carefully applied to select an optimal set from entire feature space. The performance results show that the MVMD is the key element for SCA detection performance, and the proposed algorithm is computationally efficient while featuring greater detection performance compared to previous publications.**

## I. INTRODUCTION

Recently, Internet of Things (IoT) has enabled collecting data on an unprecedented scale and gathering context-oriented information related to our physical, behavioral, and psychological health [1, 2]. Hence, applying big data analytics for mining information, extracting knowledge, and making predictions/inferences for healthcare issues has been attracting unprecedented attention [3, 4]. In this context, this paper is focused on the application of machine learning-based methods for the detection of sudden cardiac arrest.

The sudden cardiac arrest (SCA), one of the most dangerous heart diseases, is caused by the shockable (SH) rhythm including ventricular fibrillation (VF) and ventricular tachycardia (VT). Deaths are the most popular consequences of the SCA without life-saving support [4]. At present, it is well-known that the Automated External Defibrillator (AED) is considered as the most productive device for patients who are under life threating of SCA. Moreover, the most important element of AED is the shock advice algorithm (SAA), which performs quick detection to address whether SH rhythms are visible.

Lately, intelligent techniques including machine learning (ML) and deep learning (DL) have been employed to construct the SAA for the AED. Obviously, the utility of intelligent methods offers better benefit than that of conventional ones in terms of performance improvement. Furthermore, while processing massive databases is a disadvantage of conventional method, doing so lies at the heart of intelligent techniques. Such massive processing and computing is possible is thanks to the ultra-high capacity enabled by the next-generation transport networks [5-9], paving the way for wide adoption of cloud computing. Indeed, the authors of [10] propose a SAA algorithm using support vector machines (SVM) archived by the use of cross validation method. In addition, a set of input features are identified as the most informative set among a large number of input features. In [11], a set of ML algorithms and input features are investigated using statistical valid manner. The detection performance is improved significantly due to application of modified variational mode decomposition (MVMD) technique for signal processing. As a result, a final set of features are identified, which extracted from both SH and NSH signals. The VMD technique is further modified for feature extraction used as the input of boosted-cart classifier proposed as the diagnosis algorithm for the SAA [12]. Three channels, which are constructed from original ECG segments using MVMD, are fed into convolutional neural networks (CNN) in [13]. A mixed convolutional and long short-term memory model is suggested as the most effective SAA for the AED [14]. Here, a CNN plays a role of feature extraction, where a set of deep features are collected and then put into another DL algorithm.

Certainly, the SAA design using DL algorithms shows better detection performance than that using ML techniques. Furthermore, the utility of ECG channels constructed from original ECG signal and DL algorithms also contribute significantly to performance improvement of the SAA for the AED. However, the use of DL for the SAA design offers considerable challenges in comparison with ML techniques. Firstly, the DL methods require a huge amount of databases for model training, which results in a large difficulty for data collection and cooperation between researchers. Secondly, there are many parameters which need to be tuned for searching an optimal DL structure. Indeed, the parameter tunning is an exhaustive procedure and essential time-consuming. Therefore, we propose an SH/NSH rhythm detection algorithm for the SAA using a ML method in combination with MVDM for the feature extraction. The main contributions are 1) Investigation of feature extraction performance by MVMD in terms ML methods. 2) Prompt proposal of a ML model as the SAA for the AED.

In this paper, we propose a simple and effective shock advice algorithm based on AED. we first generate two different signals namely SH and NSH signals from the original ECG segment. Then, feature space includes various features extracted from three signals [11], [13] which is used as the input of the k-nearest neighbor (KNN) based feature selection algorithm for identification of the most relevant subset of features on the training data. The KNN is also employed to validate the performance of the selected subset of features on the testing data. Experimental results prove that our proposed algorithm obtains equivalent SCA detection performance to that of conventional works, i.e. given in [13] and [14] however, our algorithm is much simpler than those with respect to model construction.

## II. DATA

The Creighton University Ventricular Tachyarrhythmia Database (CUDB) and the MIT-BIH Malignant Ventricular Arrhythmia Database (VFDB), are used for development of proposed method. There are 35 8-min-records of the CUDB and 22 35-min-records of the VFDB. The signal annotations are VF, VT, non-VF, ventricular flutter, normal sinus rhythm and categorized into SH and NSH rhythms.

The frequency of 250Hz is used to sample all of records. As a result, there are 57 records of the total, which are divided into 70% and 30% as the training and testing data, respectively. All records are further separated into 8s-segment. Various steps of data processing are applied to make ECG signal better for further use. Signal smooth is generated by the five-order moving average filtering followed by the high-pass filtering for removal of baseline wander. Noise caused by high frequency is eliminated by utility of second-order low-pass Butterworth filtering.

**TABLE 1. DETAILS OF TRAINING AND TESTING DATA**

| Training data | | Testing data | | Total | |
|---|---|---|---|---|---|
| *SH segment* | *NSH segment* | *SH segment* | *NSH segment* | *SH segment* | *NSH segment* |
| 1464 | 7195 | 2273 | 7572 | 3737 | 14767 |

## III. METHOD

We use sequential forward feature selection (SFFS) in combination with KNN algorithm to select the most relevant subset of features, which are extracted from three ECG signal namely original ECG, SH, and NSH signals using MVMD techniques on the training data. Then, KNN is also adopted for validation of selected features subset on the testing data using 5-fold cross validation method. Figure 1 shows a diagram of our method.

K-Nearest Neighbors (KNN) is a machine-learning algorithm, which aims at determination of class for data samples. The method is now extended to data mining area. The major idea is to use training data including various data samples which are classified by a number of variables. Data samples are deployed on a grid and used as the neighbors of a new data sample. By computing repeatedly the distances between new data sample and given groups of K samples on a grid, the label is assigned for the new data sample if the distance is the smallest.

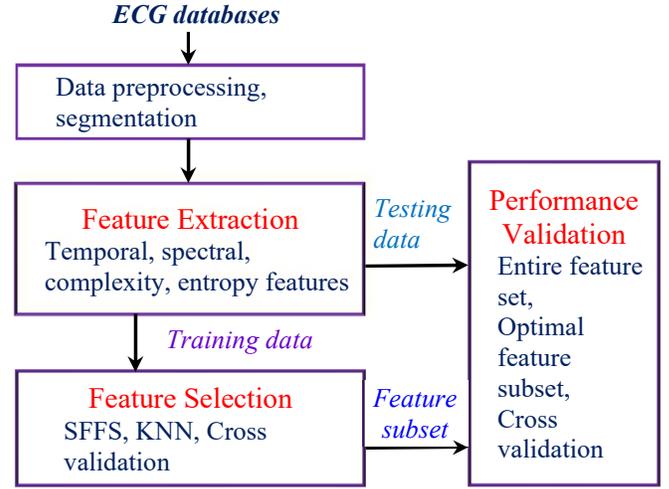

Fig 1. Method diagram

### A. Signal Generation

We generate SH and NSH signals using MVDM as proposed in [13]. The main idea is that investigation of the bandwidth for SH and NSH rhythms by the power spectrum using Fourier transform. Then, approximated bound of power spectrum for both of SH and NSH signal is addressed and used as the key element for modification of the VMD algorithm. Indeed, the frequency bound is identified as 10 Hz, for which the most effective SH and NSH signal are produced. A few modifications are implemented for the VMD to generate 10 signals so that sum of 4 signals with lower center frequencies than 10 Hz is considered as NSH signal. Sum of other signals is the SH signal due to higher center frequencies than selected bound. It is noteworthy that the signal with 0 center frequency is removed for avoidance of low frequency interference. Figure 2 and Figure 3 show the normal and abnormal ECG segments with their SH and NSH signals

### B. Feature Extraction

The feature space used for this work is suggested in [13]. The features are grouped into three categories such as temporal, spectral, complexity features, and entropy features. As a result, there are 93 features extracted from original ECG, SH, and NSH signals.

Temporal features consist of threshold crossing interval, threshold crossing sample count, mean absolute value, standard exponential algorithm, modified exponential, bCP, Count1, Count2, and Count3.

Spectral features include VF-filter leakage measure, spectral analysis (M and A2), center frequency, power spectral analysis, center power, bWT, bW, Li.

Complexity features contains phase space reconstruction, Hilbert transform, complexity measure, covariance calculation, area calculation, frequency calculation, and Kurtosis.

Entropy features are dispersion entropy, sample entropy, Energy, Renyi entropy, Fuzzy entropy, and Wavelet entropy.

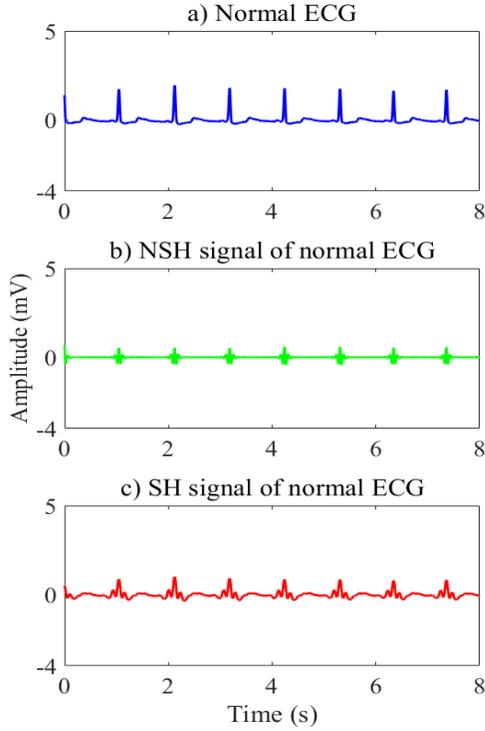

Fig 2. Normal ECG segment and its SH, NSH signals

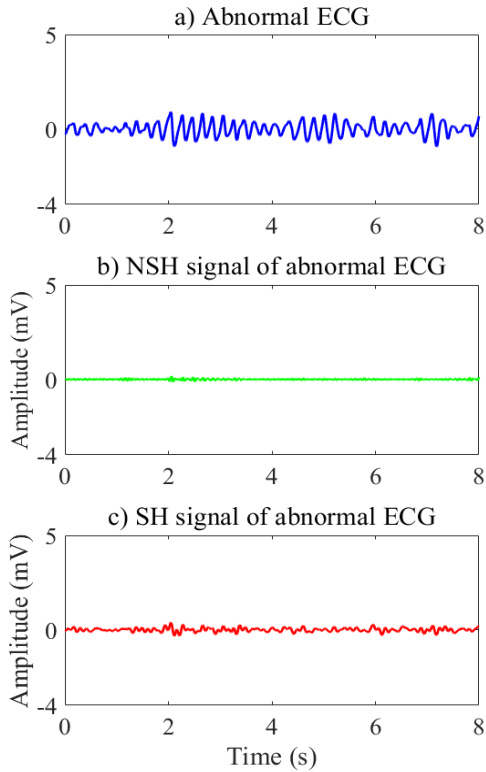

Fig 3. Abnormal ECG segment and its SH, NSH signals

### C. Feature Selection

We employ SFFF based feature selection algorithm to identify the optimal subset of features from input feature space of 93 features. At first, individual features of the input feature space are sorted by their scores from the highest to the lowest. The SFFF collect only one feature during each repetition to construct a new feature subset. Cross validation-based method is repeated 50 times for the KNN classifier using above subset of features each time. Moreover, a set of new parameters of the KNN classifier is tuned on the training data with that subset of collected features. Figure 4 shows a workflow of SFFS method.

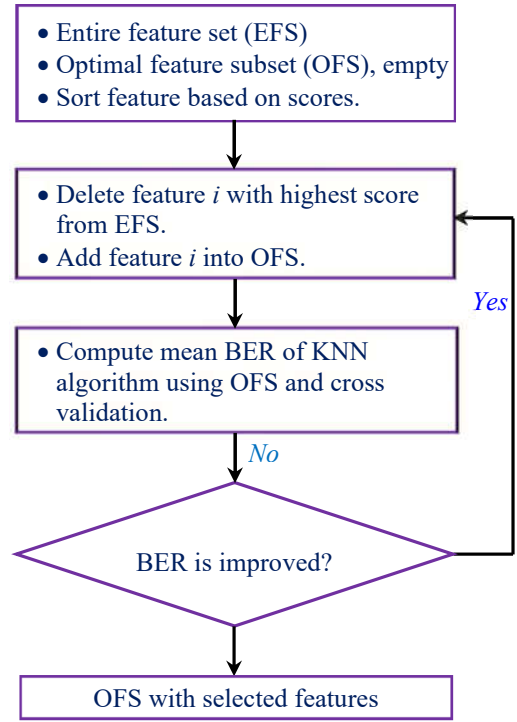

Fig 4. SFFS workflow

### D. Performance Validation

The entire feature space and optimal feature subset are put into the KNN classifier for performance validation on the testing data. Here, the 5-fold cross validation is applied to divide randomly testing data into 5 folds. A fold is for model evaluation and others are for model training. This procedure is implemented 5 times to ensure that every fold is become as evaluation data. After that, procedure is repeated with randomly division of testing data for 50 times. Mean and standard deviation of model performance are computed for further comparison and estimation of the selected feature subset and KNN model.

## IV. SIMULATION RESULTS AND DISCUSSION

We use four performance parameters for model estimation, which are accuracy (Ac), sensitivity (Se), specificity (Sp), and balanced error rate (BER). Ac measures segments classified correctly, Se and Sp show the SH and NSH segments addressed correctly. BER is computed as 1-(Se+Sp)/2.

## A. Sequential Forward Feature Selection

A subset of 36 features is selected from input feature set using SFFF method in combination with fitness function of KNN algorithm. Figure 5 show mean BER of KNN classifier using various feature subsets ranging from 1 to 93 features.

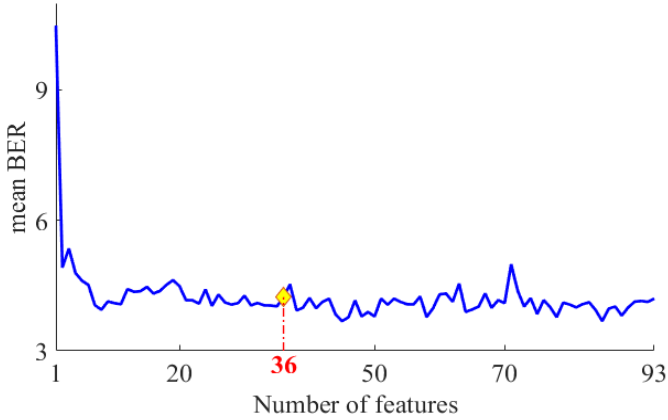

Fig 5. Mean BER of KNN classifier using individual feature subsets.

## B. Performance Estimation and Comparison

Cross validation procedure is implemented to validate performance of the KNN classifier using entire feature set and optimal feature subset as shown in Table 2. Performance of KNN classifier using optimal feature subset is higher than that using all input features.

**TABLE 2. KNN CLASSIFIER PERFORMANCE USING ENTIRE FEATURE SET AND OPTIMAL FEATURE SUBSET**

| Performance<br>Feature set | BER (%) | Ac (%) | Se (%) | Sp (%) |
|---|---|---|---|---|
| Optimal features (93) | 1.7 | 99.2 | 96.7 | 99.7 |
| Entire features (36) | 3.9 | 96.2 | 94.5 | 99.5 |

Our proposed algorithm is compared to recent publications, that are given in [13] and [14], using similar databases in terms of performance as shown in Table 3. The gaps between performance values are insignificant.

**TABLE 3. PERFORMANCE COMPARISON OF PREVIOUS WORKS AND OUR PROPOSAL**

| Ref. | Ac (%) | Se (%) | Sp (%) | Method |
|---|---|---|---|---|
| [13] | 99.3 | 97.1 | 99.2 | DL |
| [14] | 99.1 | 99.7 | 98.9 | DL |
| Our Proposal | 99.2 | 96.7 | 99.7 | ML |

To improve the performance of the SCA diagnosis, many studies have been concentrated on utility of ML and DL techniques. Compared to ML, DL offers many advantages, which are no feature extraction, feature ranking, feature selection, feature engineering. Moreover, only preprocessed data is needed for the input of DL algorithms, which automatically extract feature set through various layers. In other words, the DL proposes a self-taught learning mechanism by which the deep features are learnt layer by layer for the purpose of detection performance improvement.

However, utility of the DL techniques also encounters with few challenges that require large time and effort to overcome in terms of model construction, model selection. Indeed, one of the most difficulty is the selection of parameters for optimal DL model. Very often, an optimization algorithm is needed to search for a set of optimal parameters related to model and layer structures. As a result, this is a very time-consuming procedure and requires large resources for simulation.

From above analysis, a representative ML model namely K-nearest neighbor algorithm is considered for the SAA design applied in the AED. As shown in Table 3, the performance of this work using ML technique is similar to that of existing publications using DL methods. It means that ML or DL does not play an important role of SAA design. In contrast, the most contributive part is the MVMD, which processes, decomposes input ECG signal and then generates two different signals representing SH and NSH components of the original ECG segment.

Obviously, the NSH signal shows the higher amplitude for the normal ECG than abnormal ECG signals as shown in Figure 2.b and Figure 3.b. Hence, NSH signal is potential for extraction of features, which use for detection of normal rhythms on the ECG signal. The SH signal of abnormal ECG signal is unstable in comparison with that of normal ECG as shown in Fig 2.c and Fig 3.c. As a result, the SH signal can be used to extract features which detect abnormal rhythm from the original ECG segments.

Conventional method for SCA detection is to research for normal element from ECG signal, which shows very different waveforms for the patients who are under SCA and normal citizens. By construction of SH and NSH signal using the MVMD technique, features, which represent both SH and NSH components of original ECG segment, are extracted for the input of ML algorithm. In other words, the proposed algorithm can detect both SH and NSH rhythms on the ECG signal at the same time. This is outstanding characteristic compared to conventional methods, which diagnose only normal components.

## V. CONCLUSIONS

Correct detection of SCA is important at the first place, which allows the medical experts and emergency services to response quickly to save patient's lives.

In this paper, we proposed a SCA detection algorithm, which is applied for the AED. The algorithm is constructed with the use of MVMD technique and SFFS method in combination with cross validation procedure on the training data to produce an optimal feature subset. Moreover, the detection performance of the KNN classifier is validated carefully on testing data using cross validation. The results show that the proposed algorithm confirms similar SCA detection performance to exiting publications but simpler than those with respect to model construction. In addition, the MVMD is the most important element to improve the final detection performance in terms of SCA design.